\documentclass[12pt]{article}
\pdfoutput=1
\usepackage{amsmath,amsfonts,amssymb}
\usepackage{hyperref}
\usepackage{graphicx}
\usepackage{slashed}
\usepackage{ulem}
\usepackage{cite}
\usepackage[all]{xy}
\usepackage{array}
\usepackage[margin=.5cm]{caption}

\usepackage{color}

\newcommand \be{\begin{eqnarray}}
\newcommand \ee{\end{eqnarray}}

\makeatletter\@addtoreset{equation}{section}\makeatother

\DeclareMathOperator{\llangle}{\big\langle\hspace{-1.2mm}\big\langle\hspace{-.5mm}}
\DeclareMathOperator{\rrangle}{\hspace{-.5mm}\big\rangle\hspace{-1.2mm}\big\rangle}

\DeclareMathOperator{\tr}{tr}

\DeclareMathOperator{\Li}{Li}

\def\bF{{\mathbb{F}}}
\def\bA{{\mathbb{A}}}

\def\bR {\mathbb{R}}

\newcommand{\bea}{\begin{eqnarray}}
\newcommand{\eea}{\end{eqnarray}}
\newcommand{\beq}{\begin{equation}}
\newcommand{\eeq}{\end{equation}}
\newcommand{\bal}{\begin{equation}\begin{aligned}}
\newcommand{\eal}{\end{aligned} \end{equation}}

\newcommand{\vev}[1]{{\left< {#1} \right>}}

\newcommand{\eqn}[1]{(\ref{#1})}

\newcommand{\sect}[1]{Section~\ref{sec:#1}}

\newcommand{\address}[1]{\vbox{\center\em#1}}
\renewcommand{\title}[1]{\vbox{\center\huge{#1}}\vspace{5mm}}

\newcommand{\bD}{{\mathbb D}}

\newcommand{\cC}{{\mathcal C}}

\newcommand{\cN}{{\mathcal N}}
\newcommand{\cP}{{\mathcal P}}

\newcommand{\cO}{{\mathcal O}}
\newcommand{\cS}{{\mathcal S}}

\newcommand{\m}{m} 

\begin{document}
\bibliographystyle{utphys2}

\begin{titlepage}
\rightline{\tt NORDITA-2017-22}
\begin{center}

\vspace{20mm}

\title{The Wilson loop CFT: Insertion dimensions and structure
constants from wavy lines}

\vspace{10mm}

\renewcommand{\thefootnote}{$\alph{footnote}$}

Michael Cooke$^1$\footnote{\href{mailto:cookepm@tcd.ie}
{\tt cookepm@tcd.ie}},
Amit Dekel$^2$\footnote{\href{mailto:amit.dekel@nordita.org}
{\tt amit.dekel@nordita.org}}
and
Nadav Drukker$^1$\footnote{\href{mailto:nadav.drukker@gmail.com}
{\tt nadav.drukker@gmail.com}}
\vskip 5mm
\address{
$^1$Department of Mathematics, King's College London,
\\
The Strand, WC2R 2LS, London, United-Kingdom
}
\address{
$^2$Nordita, KTH Royal Institute of Technology and Stockholm University,\\
Roslagstullsbacken 23, SE-106 91 Stockholm, Sweden
}

\renewcommand{\thefootnote}{\arabic{footnote}}
\setcounter{footnote}{0}

\end{center}

\vspace{8mm}
\abstract{
\normalsize{
\noindent
We study operator insertions into the $1/2$ BPS Wilson loop in $\cN=4$ SYM theory
and determine their two-point coefficients, anomalous dimensions
and structure constants. The calculation is done for the first few lowest dimension
insertions and relies on known results for the expectation value of a smooth Wilson loop.
In addition to the particular coefficients that we calculate, our study elucidates the
connection between deformations of the line and operator insertions and between the
vacuum expectation value of the line and the CFT data of the insertions.
}}
\vfill

\end{titlepage}

\section{Introduction}
\label{sec:intro}

One of the most important questions in $\cN=4$ SYM theory, as in any gauge
theory is the evaluation of the vacuum expectation value (VEV)
of arbitrary Wilson loop operators.
Ideally one would like to calculate that for any value of the coupling.
Historically, weak coupling was the only feasible regime in which to work,
but holography makes the strong coupling regime accessible too.
Integrability based techniques and localization have led to arbitrary coupling
results in certain cases.

In $\cN=4$ SYM, the expectation value of a smooth Wilson loop, which is finite,
is invariant under non-singular conformal transformations.
This statement is no longer true for Wilson loops along singular curves, or for Wilson
loops with insertions of operators into them. Cusps and operator insertions may
give rise to divergences in perturbation theory and consequentially acquire
anomalous dimensions
\cite{Polyakov:1980ca,Brandt:1981kf,Craigie:1980qs, Brandt:1982gz, 
Dorn:1986dt, Polyakov:2000jg, polyakov}.
The $1/2$-BPS straight and circular Wilson loops are the
simplest candidates into which to introduce cusps and/or operator insertions. Indeed
this problem was studied in \cite{drukker,Drukker:2012de,Correa:2012hh}
and a set of boundary thermodynamic Bethe ansatz
equations that calculate their spectrum was written down.%
\footnote{In the case of the cusp, each ray is on its own $1/2$-BPS.}
These equations led to a numerical solution of the quark-antiquark potential
in this theory \cite{Gromov:2016rrp}.

In this note we study operator insertions into $1/2$-BPS Wilson loops from a different
perspective. In addition to determining some
of their anomalous dimensions, we also find
some structure constants.%
\footnote{In this note we do not consider cusps.}
To do that we employ the correspondence between small deformations of Wilson loops
and Wilson loops with operator insertions.

We define the Wilson loop with insertions as
\beq\label{op_ins}
W\left[\cO^{(1)}(x(s_1))...\cO^{(n)}(x(s_n))\right]=
\frac{1}{N}\tr\cP\left[\cO^{(1)}(x(s_1))...\cO^{(n)}(x(s_n))\,
e^{\int\left(i\dot x^\mu A_\mu(x(s)) - |\dot x| \Phi^1(x(s))\right)ds}\right].
\eeq
where%
\footnote{We work with spacetime
signature $(+---)$.
Our conventions are discussed in more detail in
Section~\ref{sec:def_str} and Section~\ref{sec:op_reps}.}
$|\dot x|\equiv\sqrt{-\eta_{\mu\nu} \dot x^\mu \dot x^\nu}$
and the integration is over the straight line.
The expectation value of \eqn{op_ins} can be viewed as the
correlation functions of the operator insertions, so we define the
VEV of the Wilson
loop to be exactly that
(henceforce we identify $s_i=x(s_i)$)
\beq\label{c_fxns}
\llangle\cO^{(1)}(s_1)...\cO^{(n)}(s_n)\rrangle=
\vev{W\left[\cO^{(1)}(s_1)...\cO^{(n)}(s_n)\right]}/\vev{W}.
\eeq
These correlation functions satisfy the axioms of a CFT.%
\footnote{More precisely, this is a defect CFT, and does not have
an energy-momentum tensor that is decoupled from the bulk,
see, \textit{e.g.}, \cite{Cardy:1984bb, McAvity:1995zd, Karch:2000gx, DeWolfe:2001pq}.}

The key to this is that the residual symmetry preserved by the $1/2$-BPS straight
line includes an $SO(1,2)$ group of conformal transformations along the line%
\footnote{The case of the circle could be analyzed just as well, and would give
essentially the same result.
However, the notation in the case of the line
is simpler, as the tangent and normal directions are constant.}.
The operator insertions can be classified by representations of this group (or
by the full $OSp(2,2|4)$ preserved by the line).
Furthermore one may define primary operators under the conformal group
(and under the superconformal group).

The correlation functions \eqn{c_fxns} are constrained in the same manner as in a usual CFT.
For example the Wilson loop with two scalar insertions satisfies the Ward identity for dilatation
\beq
iD\llangle \cO^{(1)}(s)\cO^{(2)}(0)\rrangle=
\left(\Delta_{\cO^{(1)}} + \Delta_{\cO^{(2)}}
+ s\,\partial_s\right)\llangle \cO^{(1)}(s)\cO^{(2)}(0)\rrangle
=0\,,
\eeq
which is identical in form to that of the two-point function of local operators, solved
by
\beq\label{scalar}
\llangle\cO^{(1)}(s)\cO^{(2)}(0)\rrangle=
\frac{a_{\cO^{(1)}\cO^{(2)}}(\lambda)}{s^{\Delta_{\cO^{(1)}}+\Delta_{\cO^{(2)}}}}\,,
\eeq
where $\Delta_{\cO^{(n)}}=\Delta_{\cO^{(n)}}(\lambda)$ and we associate dimensions
(classical and anomalous) to the insertions.
The two-point function of primaries vanishes unless they have the same dimension
due to special conformal symmetry.
$a_{\cO^{(1)}\cO^{(2)}}$, which below is denoted as $a_{\cO^{(1)}}$, is
the coefficient of the two-point function and is well defined, when the operators are
related to deformations of the Wilson loop, as they are in our analysis \cite{correa}.

In a similar way, the three-point function of scalar primary operators satisfies
\beq
\llangle\cO^{(1)}(s_1)\cO^{(2)}(s_2)\cO^{(3)}(s_3)\rrangle=
\frac{c_{(123)}(\lambda)}{|s_{12}|^{\Delta_1+\Delta_2-\Delta_3}
|s_{13}|^{\Delta_1-\Delta_2+\Delta_3}
|s_{23}|^{-\Delta_1+\Delta_2+\Delta_3}}\,,
\eeq
where $c_{(123)}(\lambda)$ is the structure constant and $s_{ij}=s_i-s_j$.
The four-point function is
\beq
\label{gen-4}
\llangle\cO^{(1)}(s_1)\cO^{(2)}(s_2)\cO^{(3)}(s_3)\cO^{(4)}(s_4)\rrangle=
\frac{G_{1234}(u)}
{\prod_{i<j}|s_{ij}|^{\Delta_i+\Delta_j-\Delta}}\,,
\qquad
\Delta=\frac{1}{3}\sum_{i=1}^4 \Delta_i
\,,
\eeq
where $G_{1234}$ is a function of the real cross-ratios $u$. It can in principle
be determined from the structure constants via the operator product expansion,
but we shall not explore that relation.

Above we have considered scalar primary operators. All of these
constraints generalize to descendant operators and to operators
with tensor structure. The correlation function of a descendant operator
is obtained from the correlation function of the primary operator
by the application of the lowering operator of the residual conformal algebra. For
example, for the $1/2$ BPS line along the $x^3$ direction, this lowering
operator is $\partial_3$.

The correlation functions of tensor operators are constructed from the inversion tensor
$I_{\mu\nu}(x)=\eta_{\mu\nu} - 2 \frac{x_\mu x_\nu}{x^2}$ and from the vector
$Y^\mu(x_1, x_2, x_3)=\frac{x_{13}^\mu}{x_{13}^2} -\frac{x_{23}^\mu}{x_{23}^2}$.
For insertions along a line, the former
reduces in the transverse directions to $I_{ij}=\eta_{ij}$,
with $i=0,1,2$ and the latter vanishes $Y^i=0$.
Let us consider a vector primary operator $\cO_i$ as an example.
Its two-point function is given by
\beq\label{vector}
\llangle\cO_i(s)\cO_j(0)\rrangle=
\frac{a_{\cO}(\lambda)\eta_{ij}}{s^{2\Delta_{\cO}}}\,,
\eeq
and its three-point function is constrained to vanish.
This generalizes straightforwardly to higher rank tensor operators.
In this way, as for a CFT, the correlation functions of the insertions are
determined up to a set of coefficients, the CFT data.

\begin{figure}[ht]
\centering
\includegraphics[width=.85\textwidth]{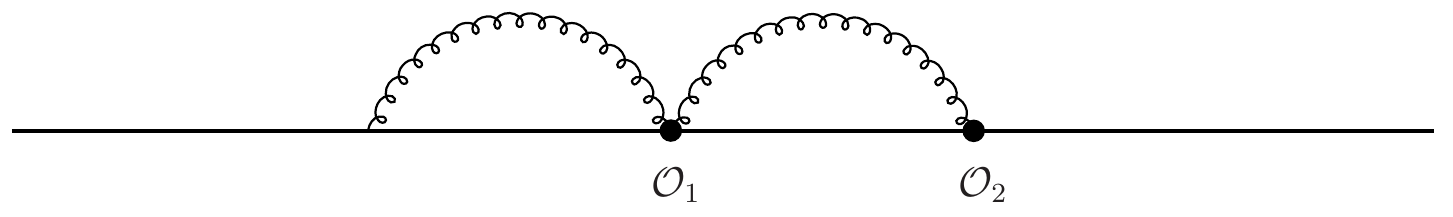}
\caption[Operator insertion Feynman diagram]
{A Feynman diagram contributing to $\llangle\cO_1\cO_2\rrangle$.
The straight line represents the $1/2$ BPS line.
The wavy lines represent propagators in the theory, the exact form of
which depends on the operators $\cO_1$ and $\cO_2$.}
\label{feyn}
\end{figure}

The anomalous dimensions can be determined in principle at all values of the gauge
coupling by using the tools of integrability. Thus far no analogous techniques were
developed to understand the three-point functions of insertions. Alternatively one can
use Feynman diagrams to evaluate any of those correlation functions.
For example, Figure~\ref{feyn} shows one Feynman diagram contributing to
the two-point function of insertions at two-loops in the coupling.
Note that the correlation function includes interactions
between the insertions
and the Wilson loop itself.

In this note we take a different approach, based on the
calculation of smooth Wilson loops which are small deformations
of the straight line. Such deformations can be written as integrals of local
operator insertions into the straight line. In the following we classify the low
dimension insertions into the straight Wilson line and find the exact mapping between
deformations and operator insertions by performing a functional Taylor expansion.

The vacuum expectation value of the nearly straight Wilson loop may be calculated
in perturbation theory and in certain instances also in string theory,
giving expressions at weak and strong coupling \cite{semenoff}.
These calculations should match the appropriate sum of integrated
$n$-point functions of the appropriate insertions.
We demonstrate this procedure and extract the first nontrivial
anomalous dimensions and structure constants of the lowest dimension insertions.

At one-loop, \emph{i.e.} order $\lambda = g^2 N$, we only have two-point functions of
operator insertions contributing to the expectation value of the Wilson loop.
More specifically, we have tree-level two-point functions of the lowest dimension
operators. At two-loops, \emph{i.e.} order $\lambda^2$, we
have two, three and (factorized) four-point functions. The three and four-point
functions are tree-level and the two-point functions include both the one-loop
corrections to the two-point functions of the lowest dimension insertions
and the tree-level correlators of composite operators. As we only have tree-level
two-point functions of composite operators, we do not expect operator mixing
at this order.

To perform the calculation,
we use known compact expressions for the planar vacuum
expectation values at one and two-loops at weak coupling.
At one-loop there is the usual well-known expression for the expectation value of a
general loop. At two-loops the expression for the expectation value of a general planar
contour was found in \cite{bassetto}. We manipulate these expressions into the form of
a sum of $n$-point functions of operator insertions, such that we may read off the
coefficients of the $n$-point functions using the mapping described above.

In \cite{semenoff} the expectation value of a general deformation of the $1/2$ BPS straight
line in $\bR^4$ was computed as an expansion in the deformation. This was done to second
order in the deformation, to two-loop order at weak coupling and at leading order at
strong coupling. This computation revealed the so-called `universality' of the
deformed line - at second order in the deformation the functional form of the expectation
value is the same at one and two-loops, and also at strong coupling.
As we outline below, in the operator insertion language the universality observed
at second order in the deformation is equivalent to the statement that a specific
operator insertion is a protected operator. The coupling dependent two-point
coefficient for this protected operator was
then understood to be related to Bremsstrahlung radiation of
an accelerated
quark \cite{correa}.

The correspondence between operator insertions into the line and deformation
of the line was also checked explicitly in \cite{semenoff} at one-loop order via a
Feynman diagram computation. In \cite{alday} a further Feynman diagram
computation was performed to find the anomalous dimension of $\Phi^1$ as
an insertion in the $1/2$ BPS line coupling to $\Phi^1$. Our analysis is consistent
with the results of \cite{semenoff} and \cite{alday}. The results also satisfy several
consistency conditions imposed by the multiplet structure of the insertions.
We find several
two-point coefficients and structure constants
which have not been previously calculated.

\section{Equivalence of operator insertions and deformations}
\label{sec:equiv}

In this section we review the equivalence between small deformations of a Wilson
loop and operator insertions along the loop. The procedure is general and holds
for deformations of any smooth loop, however starting from Section~\ref{sec:def_str}
we restrict ourselves to deformations of the $1/2$ BPS straight line.

\subsection{The operator expansion}

Let us consider a Wilson loop (or line) along a contour $\cC_0$ in $\bR^{1,3}$, defined by
$x^\mu(s)$ and let $\cC$ be a deformation of this loop given by $x^\mu(s)+\delta x^\mu(s)$.
The Wilson loop $W[\cC]$ can then be written formally by the functional Taylor expansion
\beq
W[\cC]=\sum_{n=0}^\infty \frac{1}{n!} \delta^n W[\cC_0]\,,
\eeq
where $\delta$ corresponds to taking $x^\mu(s) \to x^\mu(s) + \delta x^\mu(s)$.
We do not consider deformations of the scalar couplings, beyond the change
of the magnitude of the coupling of $\Phi^1$, which is $|\dot x|$.
This expansion is valid for any value of the coupling.

It is straightforward to take the functional variation of the Wilson loop.
Clearly the first order variation $\delta W[\cC_0]$ is equivalent to the insertion of an
operator $\cS^{(1)}(s_1)$ into $W[\cC_0]$
\beq\label{def}
\delta W[\cC_0]
=\tr\cP\oint ds_1
\cS^{(1)}(s_1) e^{i\oint ds\, \bA(s)}\,,
\eeq
where $\bA(s)$ is the connection of the undeformed loop $W[\cC_0]$ and
the integrals are along the contour $\cC_0$.

The second order variation is equivalent to two insertions of $\cS^{(1)}$ in $W[\cC_0]$ as well as the insertion of a new operator $\cS^{(2)}(s)$ into the loop,
\emph{i.e.}
\beq
\delta^2 W[\cC_0]
=\tr\cP\oint ds_1\,
\cS^{(2)}(s_1) e^{i\oint ds\, \bA(s)}
+\tr\cP\oint ds_1\,ds_2\,
\cS^{(1)}(s_1) \cS^{(1)}(s_2) e^{i\oint ds\, \bA(s)}\,.
\eeq

The $\cS^{(n)}$ are related by the recursion relation
\beq\label{eq:oprelation}
\cS^{(n+1)}
= \delta\cS^{(n)}-i\delta x^\mu [A_\mu,\cS^{(n)}]
=\left(\delta x^\mu D_\mu + \delta \dot x^\mu\frac{\partial}{\partial \dot x^\mu}\right)\cS^{(n)}\,.
\eeq

We may use the above to relate the expectation values
$\langle W[\cC_0] \rangle$ and $\langle W[\cC] \rangle$.
Using the double bracket notation \eqn{c_fxns} we have
\begin{align}
\label{S1}
\frac{\delta \langle W[\cC_0] \rangle}
{\langle W[\cC_0] \rangle}
&= \oint ds_1
\llangle \cS^{(1)}(s_1)\rrangle\,,
\\
\label{S2}
\frac{\delta^2 \langle W[\cC_0] \rangle}
{\langle W[\cC_0] \rangle}
&= \oint ds_1\llangle \cS^{(2)}(s_1)\rrangle
+\oint ds_1\,ds_2\llangle \cS^{(1)}(s_1) \cS^{(1)}(s_2) \rrangle.
\end{align}
At higher orders we find
\begin{align}
\label{S3}
\frac{\delta^3 \langle W[\cC_0] \rangle}
{\langle W[\cC_0] \rangle}
&=
\oint ds_1 \llangle \cS^{(3)}(s_1)\rrangle
+\oint ds_1\,ds_2\, 3\llangle \cS^{(1)}(s_1) \cS^{(2)}(s_2) \rrangle
\nonumber\\&\quad{}
+\oint ds_1\,ds_2\,ds_3\, \llangle \cS^{(1)}(s_1) \cS^{(1)}(s_2) \cS^{(1)}(s_3)\rrangle
\\
\frac{\delta^4 \langle W[\cC_0] \rangle}
{\langle W[\cC_0] \rangle}
&=\oint ds_1\, \llangle \cS^{(4)}(s_1)\rrangle
+\oint ds_1\,ds_2\, \left(4\llangle \cS^{(1)}(s_1) \cS^{(3)}(s_2) \rrangle
+3\llangle \cS^{(2)}(s_1) \cS^{(2)}(s_2) \rrangle\right)
\nonumber\\&\quad{}
+\oint ds_1\,ds_2\,ds_3\, 6\llangle \cS^{(1)}(s_1) \cS^{(1)}(s_2) \cS^{(2)}(s_3)\rrangle
\label{S4}
\nonumber\\&\quad{}
+\oint ds_1\,ds_2\,ds_3\,ds_4\, \llangle \cS^{(1)}(s_1) \cS^{(1)}(s_2) \cS^{(1)}(s_3)\cS^{(1)}(s_4) \rrangle.
\end{align}
We restrict our analysis to the fourth order in the deformation.

\subsection{Deformations of the straight line and explicit form of the operators}
\label{sec:def_str}

In this section, adopting the conventions of \cite{belitsky,groeger,wiegandt}, we
explain the explicit form of the operators $\cS^{(n)}$
for general deformations of the $1/2$ BPS Wilson line in $\mathbb{R}^{1,3}$
(with signature $(+---)$).%
\footnote{With this space-time signature, supersymmetry implies
also a negative inner product between the scalars, so the propagator is
$\vev{\Phi^1(x)\Phi^1(0)}=-\frac{1}{4\pi |x|^2}$.}
The $1/2$ BPS line with spacetime contour $x^\mu(s)= (0,0,0,s)$ is given by
\beq
W_{\text{BPS}}=
\tr\cP\exp\left(i\oint ds\, \bA\right),
\qquad
\bA=A_3+i\Phi^1\,.
\eeq

The first combination of operators which appears in the expansion about
a general contour is
\eqn{S1}
\beq
\cS^{(1)} =
\delta x^\mu \left(i \dot x^\nu F_{\mu \nu} -|\dot x|D_\mu \Phi^1\right)
+\frac{\delta \dot x \cdot \dot x}{|\dot x|}\Phi^1\,,
\eeq
recalling that $|\dot x|=\sqrt{-\eta_{\mu\nu}\dot x^\mu \dot x^\nu}$.
The rest of the operators are given by the relation \eqn{eq:oprelation}.
Focusing on the straight line,
by reparametrization of the loop, we can assume that the
deformations take the form $\delta x^\mu(s)= (\epsilon^i(s),0)$ where $i=0,1,2$.
$\epsilon=|\epsilon|\ll1$ serves as our expansion parameter,%
\footnote{To be more precise, we assume that each component
of $\epsilon$ is small, \emph{i.e.} $|\epsilon^i|\ll1$.}
and we assume that $\frac{d^n}{ds^n}\epsilon(s)\sim\epsilon(s)$.
To fourth order we then have
\bal\label{exp_ops}
\cS^{(1)}
&=
i \bF_{i 3}\epsilon^i\,,
\\
\cS^{(2)}
&=i D_i \bF_{j3} \epsilon^i \epsilon^j
+i F_{ij}\epsilon^i \dot\epsilon^j + \Phi^1 \dot\epsilon^2,
\\
\cS^{(3)}
&=i D_i D_j \bF_{k3}\epsilon^i \epsilon^j \epsilon^k
+ 2 i D_i F_{jk} \epsilon^i \epsilon^j \dot\epsilon^k
+ 3 D_i \Phi^1 \dot\epsilon^2 \epsilon^i,
\\
\cS^{(4)}
&=i D_i D_j D_k \bF_{m3} \epsilon^i \epsilon^j \epsilon^k \epsilon^m
+ 3 i D_i D_j F_{km} \epsilon^i \epsilon^j \epsilon^k \dot\epsilon^m
+ 6 D_i D_j \Phi^1 \dot\epsilon^2 \epsilon^i \epsilon^j
+ 3\Phi^1 \dot\epsilon^4.
\eal
Here we have introduced the notation $\bF_{i3}= F_{i3} + i D_i \Phi^1$.
Generally, the operators we encounter in this expansion are
\beq
D_{i_1}\cdots D_{i_n} \bF_{i_{n+1} 3}\,,
\qquad
D_{i_1}\cdots D_{i_n}F_{i_{n+1} j}\,,
\qquad
D_{i_1}\cdots D_{i_n}\Phi^1\,,
\eeq
with $n\geq0$.
All the indices, except for the $j$ index in the expressions above are
symmetrized, since these operators always appear contracted to $\epsilon^i$
(and $\dot\epsilon^j$). The last operator can appear with arbitrary even powers of
$\dot\epsilon$.

At strong coupling there is an algorithm to study the minimal surface associated to
small deformations of the straight Wilson line, as long as the deformations leave the contour
$\cC$ in $\bR^2$ \cite{dekel}. In that case the operator $F_{ij}$ and its derivatives do not appear in
the expansion.

\subsection{Operators representation and multiplets}
\label{sec:op_reps}

The introduction of the $1/2$ BPS space-like line breaks the $PSU(2,2|4)$
symmetry of $\cN=4$ SYM to a $OSp(2,2|4)$ residual superconformal group \cite{bianchi}.
In this section we classify the operator insertions in terms of their irreducible bosonic
representations and the supermultiplets they belong to.

The $1/2$ BPS line breaks the $SO(1,3)$ Lorentz group to $SO(1,2)$.
We take the chiral form of the gamma matrices, \emph{i.e.}
\beq
\gamma^\mu=
\begin{pmatrix}
0 & \sigma^{\mu}_{\alpha\dot\beta}\\
\bar\sigma^{\mu,\dot\alpha\beta} & 0
\end{pmatrix}\,,
\eeq
where we choose $\sigma^\mu=(1,\tau^1,\tau^3,\tau^2)$ and $\tau^a$ are the Pauli matrices.%
\footnote{And $\bar\sigma^\mu=(1,-\tau^1,-\tau^3,-\tau^2)$.}
These split according to
\beq
\sigma^{\mu,\dot\alpha\beta}
\rightarrow
\left\{\sigma^{i,\alpha\beta},\,{-i}\epsilon^{\alpha\beta}\right\}\,,
\qquad
\bar{\sigma}^\mu_{\alpha\dot\beta}\rightarrow
\left\{\sigma^i_{\alpha\beta},\,i\epsilon_{\alpha\beta}\right\}\,.
\eeq
Since we now have only one $SU(2)$ (more precisely $SO(1,2)$), there is no way to distinguish
dotted and undotted indices and indeed we can raise and lower them using $\epsilon^{\alpha\beta}$.

The $SO(6)$ R-symmetry also breaks to $SO(5)\simeq Sp(4)$.
Taking the $SO(6)$ gamma matrices to
be in a chiral form,
$\rho^{A,ab}$ and $\bar\rho^A_{ab}$,
they
split as
\beq
\rho^{I,ab}\rightarrow
\left\{\omega^{ab},\,\rho^{A,ab}\right\},
\qquad
\bar\rho^I_{ab}\rightarrow
\left\{\omega_{ab},\,\rho^A_{ab}\right\},
\eeq
with $A=2,...,6$.
Again we can use the distinguished $\rho^1=\omega$ to raise and lower indices according to
$s^a=\omega^{ab}s_b$, $s_a=s^b \omega_{ba}$
and consequently $\omega^{ab}\omega_{bc}=-\delta^a{}_c$.
Similarly, we raise and lower $SU(2)$ spinor indices as
$s^\alpha=\epsilon^{\alpha\beta}s_\beta$ and
$s_\alpha=s^\beta \epsilon_{\beta\alpha}$
with $\epsilon^{\alpha\beta}\epsilon_{\beta\gamma}=-\delta^{\alpha}{}_\gamma$.

The $OSp(2,2|4)$ Poincar\'{e} supercharges are
\beq\label{osp_sch}
Q^+_{a,\alpha}=
q_{a,\alpha}
+ \epsilon_{\alpha\beta}\omega_{ab}\bar q^{b,\beta}
=q_{a,\alpha} + \bar q_{a,\alpha}\,,
\eeq
where $q_{a,\alpha}$ and $\bar q^{a,\dot\alpha}$ are the usual $\cN=4$
SYM Poincar\'{e} supercharges. Given a superconformal primary operator,
we find their superconformal descendants by acting with $Q^+$.

We now classify the representations of the operators appearing in our
expansion and their multiplet structure.

\subsubsection{Operators of classical dimension one}

The only operator of classical dimension one in \eqn{exp_ops} is $\Phi^1$.
This is an unprotected operator, so we will find its anomalous dimension below.

The other operators in the theory of classical dimension one are $\Phi^A$, in the
$({\bf1},{\bf5})$ of $SO(1,2)\times SO(5)$. They are protected
\cite{semenoff,correa}
and do not mix
with $\Phi^1$. They do not arise in our expansion, but as we show below, some of
their descendents do.

\subsubsection{Operators of classical dimension two}

In \eqn{exp_ops} the operators
\beq\label{cd2}
\{i\bF_{i3},\quad D_i\Phi^1,\quad iF_{ij}\}\,,
\eeq
are of classical scaling dimension two.

The first two operators in \eqn{cd2} are not orthogonal,%
\footnote{Here orthogonality refers to the two-point functions.}
since $i\bF_{i3}=iF_{i3}-D_i\Phi^1$. We therefore rewrite $D_i\Phi^1$ as
a linear combination of $i\bF_{i3}$ and its orthogonal
operator $i\tilde\bF_{i3}=iF_{i3}+2D_i\Phi^1$
\beq
D_i\Phi^1=\frac{1}{3}\left(i\tilde\bF_{i3} - i\bF_{i3}\right)\,.
\eeq

$i\bF_{i3}$ is in the protected $\Phi^A$ supermultiplet, which is seen as follows.
Acting on the primary $\Phi^A$ with the supercharges \eqn{osp_sch} we find
\beq
Q^+_{a,\alpha}\Phi^A
=i\bar\rho^A_{ab}\lambda^{-,b}_\alpha\,,
\eeq
where
$\lambda^{\pm,a}_\alpha=\lambda^a_\alpha \pm \bar\lambda^a_\alpha$ are the
linear combinations of SYM fermions $\lambda^a_\alpha$ and $\bar\lambda_a^{\dot\alpha}$.
Acting again with $Q^+$ gives
\beq\label{Qlam}
Q^+_{a,\alpha}\lambda^{-,b}_\beta
=2i\delta^b{}_a \bar\sigma^i_{\alpha\beta} \bF_{i3}
+2i\epsilon_{\alpha\beta}\omega^{bc}\bar\rho^A_{ca}\bD_3\Phi^A\,,
\eeq
where the covariant derivative $\bD_3$ is defined with respect to this modified connection
$\bA$. Since the two operators in \eqn{Qlam} have different quantum numbers, we
see that $i\bF_{i3}$ is a superconformal descendant of $\Phi^A$
(and a conformal primary). The conformal descendant of $\Phi^A$ of scaling dimension two is
the second operator in \eqn{Qlam}, $\bD_3 \Phi^A$, which does not appear in \eqn{cd2}.

$iF_{ij}$ is also in the $\bf3$ of $SO(1,2)$ which is in the $\Phi^1$ supermultiplet.
Acting with $Q^+$ on $\Phi^1$ gives
\beq
Q^+_{a,\alpha}\Phi^1=
i\lambda^{+}_{a,\alpha}\,,
\eeq
and acting again with $Q^+$ gives
\beq
Q^+_{a,\alpha}\lambda^{+,b}_\beta=
2i\epsilon_{\alpha\beta}\delta^b{}_a\bD_3\Phi^1
-2\omega^{bc}\bar\rho^A_{ca}\bar\sigma^i_{\alpha\beta}D_i\Phi^A
+(\sigma^{ij})_\beta{}^\gamma F_{ij} \epsilon_{\alpha\gamma}\delta^b{}_a
+i\epsilon_{\alpha\beta}\rho^{A,bc}\bar\rho^B_{ca}\left[\Phi^A,\Phi^B\right]\,.
\eeq
Projecting onto the $(\mathbf{3},\mathbf{1})$ of $SO(1,2) \times SO(5)$ gives $F_{ij}$.

Finally $i\tilde \bF_{i3}$ is in a third supermultiplet, to which it is the superconformal primary.

The other possible insertions of classical dimension two are
\beq
(\Phi^1)^2\,,
\quad
\Phi^1\Phi^A\,,
\quad
\Phi^A\Phi^1\,,
\quad
\Phi^A\Phi^B\,,
\quad
D_i\Phi^A\,,
\quad
\bD_3\Phi^A\,,
\quad
\bD_3\Phi^1\,,
\eeq
whose quantum numbers are easy to read
($\Phi^A\Phi^B$ is comprised of the
singlet, $\bf10$ and $\bf14$ of $SO(5)$).
The three operators of dimension two appearing in the operator
expansion
\eqn{cd2}
are in the $(\mathbf{3},\mathbf{1})$ representation of $SO(1,2) \times SO(5)$.
Clearly none of the operators above are in this representation.
Thus, these operators may not mix with those appearing in the operator expansion.

\subsubsection{Operators of classical dimension three and four}

The operators of classical scaling dimension three in \eqn{exp_ops}, which arise in the expansion
are
\beq
\label{dim3-op}
\{iD_i\bF_{j3},\quad iD_iF_{jk}\}\,,
\eeq
Both of the operators above are in reducible representations of $SO(1,2) \times SO(5)$.
The former reduces to the trace $(\mathbf{1},\mathbf{1})$
and the traceless symmetric $(\mathbf{5},\mathbf{1})$.
The latter reduces to the trace $(\mathbf{3},\mathbf{1})$ and
the traceless symmetric $(\mathbf{5},\mathbf{1})$.

$iD_{\{i}\bF_{j\}3}$ is in the $\Phi^1$ supermultiplet.
Projecting onto the $(\mathbf{5},\mathbf{1})$ of $SO(1,2) \times SO(5)$, we have
\beq
\epsilon^{abcd} \bar\sigma_{\{i}^{\alpha\beta} \bar\sigma_{j\}}^{\gamma\delta} Q^+_{a,\alpha} Q^+_{b,\beta} Q^+_{c,\gamma} Q^+_{d,\delta}\Phi^1
=-1024D_{\{i}\bF_{j\}3}\,.
\eeq

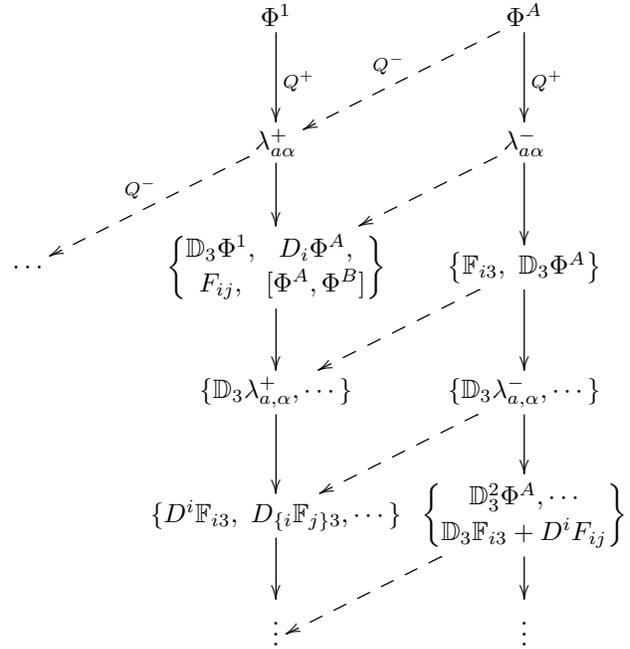
\begin{figure}[t]
\quad
\begin{tabular}{|>{$}l<{$}|>{$}l<{$}|>{$}l<{$}|}
\hline
\text{Rep}&\text{Operator} &\text{Super-}\\
&&\text{multiplet}\\ \hline&&\\[-5mm]\hline
(1,1,1)&\Phi^1 &\Phi^1 \\ \hline
(1,1,5)&\Phi^A &\Phi^A \\ \hline
&\lambda^-_{a\alpha} &\Phi^A \\ \cline{2-3}
&&\\[-8mm](\frac{3}{2},2,4)&&\\[-3mm]
&\lambda^+_{a\alpha} &\Phi^1\\ \hline
(2,1,1)&\bD_{3}\Phi^1 &\Phi^1\\ \hline
(2,1,5)&\bD_{3}\Phi^A &\Phi^A\\ \hline
(2,1,10)&[\Phi^A,\Phi^B] &\Phi^1\\ \hline
&\bF_{i3} &\Phi^A \\ \cline{2-3}
(2,3,1)&F_{ij} &\Phi^1 \\ \cline{2-3}
&&\\[-5mm]
&\tilde\bF_{i3} &\tilde\bF \\ \hline
(2,3,5)&D_{i}\Phi^A &\Phi^1\\ \hline
&\bD_3\lambda^-_{a\alpha} &\Phi^A \\ \cline{2-3}
&&\\[-8mm](\frac{5}{2},2,4)&&\\[-3mm]
&\bD_3\lambda^+_{a\alpha} &\Phi^1\\ \hline
(3,1,1)&D^i\bF_{i3}+\cdots&\Phi^1\\ \hline
(3,1,5)&\bD_3\bD_3\Phi^A&\Phi^A\\ \hline
&\bD_3\bF_{i3}&\Phi^A\\ \cline{2-3}
&&\\[-8mm](3,3,1)&&\\[-3mm]
&D^iF_{ij}
+\frac{2}{3}\bD_3\bF_{j3}&\tilde\bF\\&+\cdots&\\ \hline
(3,5,1)&D_{\{i}\bF_{j\}3}&\Phi^1\\ \hline
\end{tabular}
\ \parbox{3.5in}{
\footnotesize
$$
\xymatrix@C=3.3cm@R=1.65cm@!0{
&\Phi^1\ar[d]^{Q^+}&\Phi^A\ar[d]^{Q^+}\ar@{-->}@[gray][dl]_{Q^-}\\
&\lambda^+_{a\alpha}\ar[d]\ar@{-->}@[gray][dl]_{Q^-}&\lambda^-_{a\alpha}\ar[d]\ar@{-->}@[gray][dl]\\
\cdots &{\begin{Bmatrix}
\mathbb{D}_3 \Phi^1, & D_i \Phi^A, \\ F_{ij}, & [\Phi^A,\Phi^B]
\end{Bmatrix}} \ar[d]
&\left\{\mathbb{F}_{i 3},~\mathbb{D}_3 \Phi^A\right\} \ar[d]\ar@{-->}@[gray][dl]\\
{} &\{\mathbb{D}_3 \lambda^+_{a,\alpha},\cdots\} \ar[d]&\{\mathbb{D}_3 \lambda^-_{a,\alpha},\cdots \} \ar[d]\ar@{-->}@[gray][dl]\\
{} & \{ D^i \mathbb{F}_{i3},~ D_{\{i}\mathbb{F}_{j\}3},\cdots\} \ar[d]&
{\begin{Bmatrix}
\mathbb{D}_3^2\Phi^A, \cdots\\\mathbb{D}_3 \mathbb{F}_{i 3}+ D^i F_{ij}
\end{Bmatrix}}
\ar[d]\ar@{-->}@[gray][dl]\\
{} & \vdots & \vdots}
$$}
\caption{The insertions of lowest dimensions with their quantum numbers
under $SO(1,2)^2\times Sp(4)$ (the first is the dimension).
The diagram shows the supermultiplets starting with the primary fields $\Phi^1$
and $\Phi^A$ and acting with the unbroken supercharges
$Q^+_{\alpha,a}$ to generate the super-descendants.
The dashed arrows show the action of the broken supercharges
$Q^-_{\alpha,a}$, which take us from one supermultiplet to another.}
\label{dg:diag2}
\normalsize
\end{figure}

The three other operators are neither primaries nor descendants. The new ingredient
that arises for operators of dimension three is mixing with fermion bilinears and indeed
the two traces of the operators in \eqn{dim3-op} mix with fermions to form
(super-)descendants of $\Phi^1$ and $\Phi^A$.

Consider the equations of motion
\bal\label{desc_i}
iD^i\bF_{i3}
&=
D^3D_3\Phi^1
+\left[D_3\Phi^1,\Phi^1\right]
+\left[D_3\Phi^A,\Phi^A\right]
+\left[\left[\Phi^1,\Phi^A\right],\Phi^A\right]
+\frac{1}{2}\left[\lambda^+_{a,\alpha},\lambda^{-a,\alpha}\right]
\\&=-\bD_3\bD_3\Phi^1
+\left[\bD_3\Phi^A,\Phi^A\right]
+\frac{1}{2}\left[\lambda^+_{a,\alpha},\lambda^{-a,\alpha}\right],
\eal
which shows that
$iD^i\bF_{i3}$
mixes with fermion bilinears (and
the other commutator) to form a descendant of $\Phi^1$. We have not analyzed
the full set of operators of this dimension to find the other
combinations orthogonal to
$\bD_3\bD_3\Phi^1$. In any case, for our
two-loop analysis
below, the mixing with the bilinears is suppressed, so to this order
it is effectively a descendant.

For $iD^iF_{ij}$, we see from the equations of motion that
\beq
iD^iF_{ij}=-\frac{1}{3}\bD_{3}\left(2i\bF_{j3} + i\tilde\bF_{j3}\right)
+\left[\Phi^1,i\bF_{j3}\right]
+\left[D_j\Phi^A,\Phi^A\right]
+\frac{1}{2}\sigma^{\alpha\beta}_j\left[\lambda^+_{a,\alpha},\lambda^{a,-}_\beta\right]\,,
\eeq
so it is an admixture of descendants in two different supermultiplets.
The final operator $iD_{\{i}F_{j\}k}$ is in neither
the $\Phi^A$ nor the $\Phi^1$ supermultiplet and does not appear in the
operator expansion.

The only operators of dimension four
which contribute to the expectation value of the deformed line
at $\cO(\epsilon^4)$ (see \eqn{complete} below) are
\beq
\{iD^n D_{(n} \bF_{m)3} \eta_{jk}
+
i D^n D_{(n} \bF_{k)3} \eta_{jm}
-\frac{2}{3} iD^n D_{(n} \bF_{j)3} \eta_{km}\,,
\quad
iD_j D^n \bF_{n3} \eta_{km}\}\,.
\eeq
Up to mixing with bilinears, these are descendants of $i\bF_{i3}$ \emph{c.f.} \eqn{desc_i}.

\section{Extracting the CFT data}
\label{sec:extract}

We now use the equivalence between the operator insertions and small
deformations of the 1/2 BPS Wilson line to extract the anomalous dimensions
and two and three-point function coefficients at weak coupling.
The idea is to use the known perturbative expressions for a general Wilson
loop at one and two-loops, expand these expressions
in the deformation parameter $\epsilon$ and
rewrite
them in the form of
integrated correlation functions of insertions.

At one-loop the Wilson loop is sensitive to two-point functions only.
At two-loops things get more interesting, the Wilson loop is now sensitive to two,
three and four-point functions.
Beyond this simple statement, which is evident from the number of integrals
in the perturbative expressions below \eqn{gen_1l}, \eqn{gen_2l}, there is a
relation between the loop order in the Wilson loop evaluation and the CFT data.
The one-loop expansion of the Wilson loop is only sensitive to the \emph{classical}
dimension of the operators and their \emph{classical} two-point function coefficients.
The two-loop expansion of the Wilson loop supplies the \emph{one-loop}
anomalous dimension and coefficients and the \emph{classical} structure
constant. The four-point functions at this loop order is factorized and supplies
no new information, just the classical dimensions already found before.

A higher loop analysis using this method would require expressions for generic
Wilson loops, which are not available at present.
It should be possible using our present technology to go to higher orders in
$\epsilon$, though issues of mixing of operators could arise.

We comment on the feasibility of extracting information from
the holographic dual at strong coupling in Section~\ref{sec:discuss}.

\subsection{The one and two-loop expectation values for general contours}

At one-loop, the expectation value of a Wilson loop along a
general smooth contour is given by the well known expression
which with the mostly negative signature is
\beq\label{gen_1l}
\langle W[\cC] \rangle_{\text{1-loop}}
=
-\frac{\lambda}{16\pi^2}\oint ds_1\,ds_2\,
I(s_1,s_2)\,,
\qquad
I(s_1,s_2)
=\frac{\dot x_1\cdot \dot x_2 +|\dot x_1| |\dot x_2|}{x^2_{12}}\,.
\eeq

At two-loops there is an analogous expression for a general smooth
contour with a constant scalar coupling.
The sum of all two-loop Feynman diagrams in the planar approximation
can be combined to the elegant expression \cite{bassetto}
\bal
\label{gen_2l}
\langle W[\cC] \rangle_{\text{2-loop}}=
&-\frac{\lambda^2}{128\pi^4}\oint ds_1\,ds_2\,ds_3\,
\epsilon(s_1,s_2,s_3) I(s_1,s_3) \frac{x_{32}\cdot \dot x_2}{x_{32}^2}\log\frac{x_{21}^2}{x_{31}^2}
\\&\hskip-2cm{}
+\frac{\lambda^2}{2}\left(\frac{1}{16\pi^2}\oint ds_1\,ds_2 \, I(s_1,s_2)\right)^2
-\frac{\lambda^2}{64\pi^4}\int_{s_1>s_2>s_3>s_4} ds_1\,ds_2\,ds_3\,ds_4\,
I(s_1,s_3)I(s_2,s_4)\,.
\eal
where $\epsilon(s_1,s_2,s_3)$ is completely antisymmetric and takes the
value $1$ for $s_1>s_2>s_3$.

\subsubsection{Divergences and regularization}
\label{sec:divs}

The one and two-loop integrals above are finite. However if we split the one-loop
integral as
\beq
\label{one-loop-split}
-\frac{\lambda}{8\pi^2}\int_{s_1>s_2} ds_1\,ds_2\,
\frac{\dot x_1\cdot \dot x_2 + |\dot x_1| |\dot x_2|}{x^2_{12}}
=-\frac{\lambda}{8\pi^2}\left(
\int_{s_1>s_2} ds_1\,ds_2\,\frac{\dot x_1\cdot \dot x_2}{x^2_{12}}
+\int_{s_1>s_2} ds_1\,ds_2\,\frac{|\dot x_1| |\dot x_2|}{x^2_{12}}
\right),
\eeq
then each integral is separately divergent. The same is true for the two-loop integrals.

This is indeed the procedure we follow, regularizing each of the integrals independently and
rewriting it in the form of integrated $n$-point functions. Within these final expressions we
identify the regularization independent quantities, like the anomalous dimensions.
For example, the last term in \eqn{one-loop-split} can
be integrated twice by parts to give a denominator of $x_{12}^4$, which corresponds
to the two-point function of an operator of dimension two.

We use two different regularization prescriptions. The first, point-splitting, puts a hard cutoff
on the integrals above, such that the range of integration is $s_1>s_2+\mu$, and likewise
cutoffs at infinity. At two-loops the triple and quadruple integrals
involve more cutoffs.
Boundary terms expanded around the cutoff lead to many regularization dependant terms
and overall rather messy expressions.

The other regularization we employ adds ``mass terms'', or $i\varepsilon$ terms to the
denominators, so $x_{12}\to x_{12}+i\zeta$. The resulting expressions are a bit cleaner and
involve fewer regularization dependant terms.

\subsection{Order $\epsilon^2$}
\label{sec:ep2}

At order $\cO(\epsilon)$, equation \eqn{S1} includes only the one-point function
of $i\bF_{i3}$. This vanishes as discussed in the introduction, and any divergences
that may appear are safely removed.
We therefore start our discussion at order $\epsilon^2$, also known as the wavy
line approximation, which is well studied and somewhat special \cite{semenoff, correa}.
We now rederive the known results to illustrate our approach.

Denoting by $\langle W[\cC] \rangle\big|_{\epsilon^2}$ the contributions of order
$\cO(\epsilon^2)$ to the
expectation value $\langle W[\cC] \rangle$,
the relevant terms from the expansion \eqn{S1}, \eqn{S2} with the explicit operators
\eqn{exp_ops} are
\bal\label{irreps_exp}
\langle W[\cC] \rangle\big|_{\epsilon^2}
&=\frac{1}{2!}\oint ds_1\,\left(
\epsilon_1^i \epsilon_1^j \llangle i D_{\{i} \bF_{j\}3}(s_1)\rrangle
-\frac{1}{3}\delta_{ij}\epsilon_1^i \epsilon_1^j \llangle i D^i\bF_{i3}(s_1)\rrangle\right)
\\&\quad{}
+\frac{1}{2!}\oint ds_1\,ds_2\,
\epsilon_1^i \epsilon_2^j \llangle i\bF_{i3}(s_1) i\bF_{j3}(s_2) \rrangle\,,
\eal
where $\epsilon^i_1=\epsilon^i(s_1)$ and similarly for $s_2$.
The two-point functions are constrained (since $i\bF_{i3}$ is a conformal primary)
to take the form
\beq\label{sec_con}
\langle W[\cC] \rangle\big|_{\epsilon^2}
=\int_{s_1>s_2} ds_1\,ds_2\,
\frac{a_{F}(\lambda)}{s_{12}^{2\Delta_{F}(\lambda)}} \eta_{ij} \epsilon_1^i \epsilon_2^j\,,
\eeq
where $a_{F}(\lambda)$ is the two-point coefficient of $i\bF_{i3}$ and
$\Delta_{F}(\lambda)$ is its scaling dimension. \eqn{sec_con} is true for
all values of the coupling.

\subsubsection{One-loop}
\label{sec:one-loop}

To match with the perturbative expansion of the Wilson loop, we
expand the all-coupling expression \eqn{sec_con} at weak coupling.
The $\cO(\lambda)$ term is
\beq\label{ex_opins}
\langle W[\cC]\rangle_{\text{1-loop}}\big|_{\epsilon^2}
=\lambda\int_{s_1>s_2} ds_1\,ds_2\,
\frac{a^0_{F}}{s_{12}^{2\Delta^0_{F}}} \eta_{ij} \epsilon_1^i \epsilon_2^j\,,
\eeq
where $a^0_{F}$ is the tree-level two-point function coefficient and
$\Delta^0_{F}$ is the classical scaling dimension. In this case clearly
$\Delta^0_{F}=2$, and it thus remains to determine $a^0_{F}$.

To do this, we compare the above expression with \eqn{gen_1l} for the contour $\cC$ defined by
$x^\mu=(\epsilon^i(s),s)$ and expand to order $\cO(\epsilon^2)$
\bal\label{unreg}
&\langle W[\cC] \rangle_{\text{1-loop}}\big|_{\epsilon^2}=
-\frac{\lambda}{16\pi^2}
\int^\infty_{-\infty} ds_1 \int^{s_1}_{-\infty} ds_2\,
\frac{(\dot\epsilon_1 - \dot\epsilon_2)^2}{s_{12}^2}\,.
\eal
We now split this integral into a sum of three integrals, each of which is divergent, so we need to
introduce a regularization scheme. With a cutoff parameter $\mu$ we have
\bal
\langle W[\cC] \rangle_{\text{1-loop}}\big|_{\epsilon^2}=&\,
-\frac{\lambda}{16\pi^2}
\left(\int^\infty_{-\infty} ds_1 \int^{s_1-\mu}_{-\infty} ds_2\,
\frac{\dot\epsilon_1^2}{s_{12}^2}
-2\int^\infty_{-\infty} ds_1 \int^{s_1-\mu}_{-\infty} ds_2\,
\frac{\dot\epsilon_1\cdot\dot\epsilon_2}{s_{12}^2}\right.\\
&\hspace{20mm}\left.
+\int^\infty_{-\infty} ds_1 \int^{s_1-\mu}_{-\infty} ds_2\,
\frac{\dot\epsilon_2^2}{s_{12}^2}\right)
\,,
\eal
The first and last terms are symmetric with respect to $s_1\leftrightarrow s_2$.
We may perform the integration in $s_2$ and $s_1$ for these, respectively,
without any knowledge of $\epsilon^i(s)$.
We also integrate the second term by parts to find
\bal
\langle W[\cC] \rangle_{\text{1-loop}}\big|_{\epsilon^2}
&=
-\frac{3\lambda}{4\pi^2}
\int^\infty_{-\infty} ds_1 \int^{s_1-\mu}_{-\infty} ds_2\,
\frac{\epsilon_1\cdot\epsilon_2}{s_{12}^4}
-\frac{\lambda}{8\pi^2}\int^\infty_{-\infty}ds_1\frac{\dot\epsilon_1^2}{\mu}\\
&\quad{}+\frac{\lambda}{8\pi^2}\int^\infty_{-\infty}ds_1\,
\frac{\dot\epsilon(s_1)\cdot\epsilon(s_1-\mu)}{\mu^2}
+\frac{\lambda}{4\pi^2}\int^\infty_{-\infty}ds_1\,
\frac{\epsilon(s_1+\mu)\cdot\epsilon(s_1)}{\mu^3}\,.
\eal
The terms with single integrals are all one-point functions that can be discarded and we are left with
\beq\label{mu}
\langle W[\cC] \rangle_{\text{1-loop}}\big|_{\epsilon^2}
=-\frac{3\lambda}{4\pi^2}
\int^\infty_{-\infty} ds_1 \int^{s_1-\mu}_{-\infty} ds_2\,
\frac{\epsilon_1\cdot\epsilon_2}{s_{12}^4}\,.
\eeq

We get the same result by employing mass regularization replacing $s_{12}$ in \eqn{unreg}
with $s_{12}+i\zeta_{12}$ giving
\bal
\langle W[\cC] \rangle_{\text{1-loop}}\big|_{\epsilon^2}=&\,
-\frac{\lambda}{16\pi^2}
\left(\int^\infty_{-\infty} ds_1 \int^{s_1}_{-\infty} ds_2\,\frac{\dot\epsilon_1^2}{(s_{12}+i\zeta_{12})^2}
-2\int^\infty_{-\infty} ds_1 \int^{s_1}_{-\infty} ds_2\,
\frac{\dot\epsilon_1\cdot\dot\epsilon_2}{(s_{12}+i\zeta_{12})^2}\right.\\
&\hspace{20mm}\left.+\int^\infty_{-\infty} ds_1 \int^{s_1}_{-\infty} ds_2\,
\frac{\dot\epsilon_2^2}{(s_{12}+i\zeta_{12})^2}\right)
\,.
\eal
Again we can integrate the first and last terms with respect to $s_2$ and $s_3$ respectively and integrate
the second one twice by parts to find
\beq
\label{mass_1l}
\langle W[\cC] \rangle_{\text{1-loop}}\big|_{\epsilon^2}=
\frac{i\lambda}{16\pi^2}
\oint ds_1\,
\left(\frac{\dot\epsilon_1^2}{\zeta_{12}}
+\frac{4\epsilon_1^2}{\zeta_{12}^3}\right)
-\frac{3\lambda}{4\pi^2}\int_{s_1>s_2}ds_1\,ds_2\,
\frac{\epsilon_1\cdot\epsilon_2}{(s_{12}+i\zeta_{12})^4}\,,
\eeq
where we have thrown away a total derivative. Discarding the single integral terms,
reproduces \eqn{mu}.

Comparing this with the corresponding terms in \eqn{irreps_exp}, this is the tree-level
contribution of the two-point of $i\bF_{i3}$, from which we may extract the tree-level
two-point function coefficient
\beq\label{tree}
a^0_{F}=-\frac{3}{4\pi^2}\,.
\eeq
This agrees with the expression found in \cite{semenoff}, up to an overall sign, which is due to
our choice of signature.
The terms which were discarded are regularization artifacts which contain
no physical information.

\subsubsection{Two-loop}

At order $\cO(\lambda^2)$, equation \eqn{sec_con} is
\beq\label{ex_opins_2}
\langle W[\cC] \rangle_{\text{2-loop}} \big|_{\epsilon^2}
=\lambda^2\int_{s_1>s_2} ds_1\,ds_2\,
\frac{a^1_{F} - 2a^0_F\gamma_{F}\log \frac{s_{12}}{\m}}{s_{12}^{2\Delta^0_{F}}}
\eta_{ij} \epsilon_1^i \epsilon_2^j\,,
\eeq
where $a^1_{F}$ is the one-loop two-point function coefficient, $\gamma_{F}$
is the anomalous dimension and $\m$ is a scale parameter.
Clearly if the anomalous dimension is nonzero, then $a_F^1$ can be modified
by changing $\m$, and therefore is scheme dependent. As shown below,
this is not the case for this operator, but is in fact true for most of the other
operators we encounter.
Expanding \eqn{gen_2l} to second order in $\epsilon$ gives
\beq
\label{ep2-2loop}
\langle W[\cC] \rangle_{\text{2-loop}}\big|_{\epsilon^2}
=
\frac{\lambda^2}{256\pi^4}\oint ds_1\,ds_2\,ds_3\,
\epsilon(s_1,s_2,s_3)\frac{(\dot\epsilon_1 - \dot\epsilon_3)^2
\log\frac{s_{12}^2}{s_{13}^2}}{s_{13}^2 s_{23}}\,.
\eeq

We restrict the domain of integration to $s_1>s_2>s_3$ and symmetrize the integrand accordingly.
We also introduce two independent cutoffs $\mu_1$ and $\mu_2$
\bal
\label{ep2-split}
\langle W[\cC] \rangle_{\text{2-loop}}|_{\epsilon^2}
=\frac{\lambda^2}{256\pi^4}&\Bigg(\int^\infty_{-\infty} ds_1
\int^{s_1-\mu_1}_{-\infty} ds_2 \int^{s_2-\mu_2}_{-\infty}ds_3\,
\frac{\dot\epsilon_1^2\log\frac{s_{12}^2}{s_{13}^2}}{s_{13}^2 s_{23}}\\
&\quad{}+\int^\infty_{-\infty} ds_1 \int^{s_1-\mu_1}_{-\infty} ds_2 \int^{s_2-\mu_2}_{-\infty}ds_3\,
\frac{\dot\epsilon_3^2\log\frac{s_{12}^2}{s_{13}^2}}{s_{13}^2 s_{23}}\\
&\quad{}-2\int^\infty_{-\infty} ds_1 \int^{s_1-\mu_1}_{-\infty} ds_2 \int^{s_2-\mu_2}_{-\infty}ds_3\,
\frac{\dot\epsilon_1\cdot\dot\epsilon_3\log\frac{s_{12}^2}{s_{13}^2}}{s_{13}^2 s_{23}}\Bigg)\\
&\quad{}+(\text{symmetrization})\,.
\eal
Similarly to the one-loop case, we integrate the first
term
with respect to $s_2$ and $s_3$
and the second
term
with respect to $s_1$ and $s_2$, leading to divergent one-point function
expressions which can be omitted.
For the third term we may perform the $s_2$ integral to get
\beq\label{2loop_mu}
-\frac{\lambda^2}{64\pi^4}\int^\infty_{-\infty} ds_1 \int^{s_1-\mu_1-\mu_2}_{-\infty}ds_3\,
\frac{\dot\epsilon_1\cdot\dot\epsilon_3}{s_{13}^2}
\left(\Li_2\left(\frac{\mu_2}{s_{13}}\right)-\Li_2\left(1-\frac{\mu_1}{s_{13}}\right)\right)
\eeq
We
find
here two terms, one dependent on $\mu_2$ and the other on $\mu_1$. Focusing on the first,
it is possible to shift $s_1$ by $\mu_2$ such that $\mu_2$ appears only in the integrand and $\mu_1$
only in the range of $s_3$. This clear separation of the two cutoffs simplifies the calculation of the
divergent terms and is the procedure we employ for subsequent multi-integrals in the following.

For the case at hand, these procedures are not strictly necessary, as the integrand has a simple
limit for $\mu_1,\mu_2\to0$, namely $\Li_2(0)=0$ and $\Li_2(1)=\frac{\pi^2}{6}$. One then
integrates by parts with respect to both $s_1$ and $s_3$, and the
boundary terms can be safely ignored.%
\footnote{As presented they are all divergent, but if we keep subleading terms in the expansions
of $\Li_2$, there may also be finite terms. Still those are one-point functions which should vanish, and
indeed we find that they are all total derivatives which can be integrated away.}
We find
\beq
-\frac{\lambda^2}{64\pi^4}\int_{s_1<s_3} ds_1 \,ds_3\,
\epsilon_1\cdot\epsilon_3\,\frac{d^2}{ds_1\,ds_3}\frac{-\pi^2}{6s_{13}^2}
=-\frac{\lambda^2}{64\pi^2}\int_{s_1<s_3} ds_1 \,ds_3\,
\frac{\epsilon_1\cdot\epsilon_3}{s_{13}^4}\,.
\eeq
Under the symmetrization there is another term with $\dot\epsilon_1\cdot\dot\epsilon_3$, so
again we can do the $s_2$ integral, which ends up giving an identical contribution.
Then there are four other terms with $s_2$ exchanged for $s_1$ or $s_3$, whose
contribution is minus the above. All together we find
\beq\label{fin_mu1}
\langle W[\cC] \rangle_{\text{2-loop}}|_{\epsilon^2}
=\frac{\lambda^2}{32\pi^2}\int_{s1>s_2}ds_1 \,ds_2\,
\frac{\epsilon_1\cdot\epsilon_2}{s_{12}^4}\,.
\eeq

We can also perform the integrals in \eqn{ep2-2loop} in the other regularization scheme, where
it now takes the form
\beq\label{mass_2l}
\langle W[\cC] \rangle_{\text{2-loop}}|_{\epsilon^2}
=\frac{\lambda^2}{256\pi^4}
\int^\infty_{-\infty} ds_1 \int^{s_1}_{-\infty} ds_2 \int^{s_2}_{-\infty}ds_3\,
\frac{(\dot\epsilon_1-\dot\epsilon_3)^2\log\frac{(s_{12}+i\zeta_{12})^2}{(s_{13}+i\zeta_{13})^2}}
{(s_{13}+i\zeta_{13})^2 (s_{23}+i\zeta_{23})}
+(\text{sym})\,.
\eeq
Splitting as above, the integrals of the terms analogous to the first two lines of \eqn{ep2-split}
give poles in the $\zeta_{ij}$ parameters and the third line gives again an expression which reduces
to \eqn{2loop_mu} in the $\zeta_{13}\to0$ limit (and one may also take this limit after the integration
by parts), so leading to the same result.

Comparing with \eqn{ex_opins_2} gives
\beq
a^1_{F}=\frac{1}{32\pi^2}\,,
\qquad
\gamma_{F}=0\,,
\eeq
so
\beq
\label{aF1}
a_F=-\frac{3\lambda}{4\pi^2}+\frac{\lambda^2}{32\pi^2}+O(\lambda^3)\,.
\eeq
This matches the results of \cite{semenoff} up to the difference in sign mentioned above
and is consistent with the fact that $i\bF_{i3}$ is a protected operator with vanishing
anomalous dimension. This is also the origin of the ``universality'' property, where
at all values of the coupling
\beq
\langle W[\cC] \rangle\big|_{\epsilon^2}=
\oint ds_1\,ds_2\,
\frac{a_F(\lambda)\eta_{ij}\epsilon^i\epsilon^j}{s_{12}^{4}}\,,
\eeq
and the only dependence on the coupling is in $a_F(\lambda)$. And
indeed this also can be derived at strong coupling from the $AdS$ dual, where
$a_F=-3\sqrt\lambda/\pi^2$ \cite{semenoff}.
The exact result is given by $a_F = -12 B(\lambda)$, where
$B(\lambda) = \frac{\sqrt{\lambda}I_2(\sqrt{\lambda})}{4\pi^2 I_1(\sqrt{\lambda})}$
is the Bremsstrahlung function \cite{correa}
($I_n(x)$ are modified Bessel functions).

\subsection{Order $\epsilon^3$}

At $\cO(\epsilon^3)$ we have the expression in \eqn{S3} with the explicit operators
\eqn{exp_ops} which depends on the following two and three-point functions
\bal
&\llangle i\bF_{i3}(s_1) iD^j\bF_{j3}(s_2) \rrangle,
\quad
\llangle i\bF_{i3}(s_1) iD_{\{j\bF_k\}3}(s_2) \rrangle,
\quad
\llangle i\bF_{i3}(s_1) iF_{jk}(s_2) \rrangle,
\\
&\llangle i\bF_{i3}(s_1) \Phi^1(s_2) \rrangle,
\quad
\llangle i\bF_{i3}(s_1) i\bF_{j3}(s_2) i\bF_{k3}(s_3) \rrangle.
\eal
Comparing with the table of operators in Figure~\ref{dg:diag2},
we see that all two-point functions involve pairs of operators from different supermultiplets,
and we would therefore expect them to cancel. The last expression, which is the
three-point function of $\bF_{i3}$ also vanishes, because the only invariant 3-tensor is antisymmetric.

Indeed the perturbative expressions at one-loop \eqn{gen_1l} and two-loop \eqn{gen_2l} do not
contain any terms at $\cO(\epsilon^3)$.

\subsection{Order $\epsilon^4$}

At $\cO(\epsilon^4)$ there are contributions from two, three and four-point
functions. The operator expansion at this order is quite long and messy.
As such we deal with the four, three and two-point functions separately.
We further
analyze
the two-point functions of the different multiplets separately.

\subsubsection{The four-point functions}

The sole four-point function appearing in the operator expansion \eqn{S4}
at $\cO(\epsilon^4)$ is
\beq
\label{4-insertions}
\int_{s_1>s_2>s_3>s_4}ds_1\,ds_2\,ds_3\,ds_4\,
\llangle i\bF_{i3}(s_1) i\bF_{j3}(s_2) i\bF_{k3}(s_3) i\bF_{m3}(s_4)\rrangle
\epsilon_1^i \epsilon_2^j \epsilon_3^k \epsilon_4^m\,.
\eeq

At two-loops we have the quadruple integrals on the second line of \eqn{gen_2l},
which involve two of the one-loop generalized propagator \eqn{gen_1l} and the
dependence on the deformation is
$\dot\epsilon_1^i\dot\epsilon_2^j\dot\epsilon_3^k\dot\epsilon_4^m$.
As in the one-loop loop calculation in Section~\ref{sec:one-loop}, we can perform
multiple integration by parts, to reproduce the deformation dependence as in
\eqn{4-insertions}, with the two-point functions as calculated in Section~\ref{sec:one-loop}.
\beq
-\lambda^2\int_{s_1>s_2>s_3>s_4}ds_1\,ds_2\,ds_3\,ds_4\,
\frac{a^0_{F}}{s_{13}^{2\Delta^0_{F}}} \frac{a^0_{F}}{s_{24}^{2\Delta^0_{F}}}
(\epsilon_1\cdot\epsilon_3)(\epsilon_2\cdot\epsilon_4)
+\frac{\lambda^2}{2}\left(\oint ds_1\,ds_2\,\frac{a^0_{F}}{s_{13}^{2\Delta^0_{F}}} \epsilon_1\cdot\epsilon_2\right)^2\,.
\eeq
This is exactly
the factorized form of the four-point function in
\eqn{4-insertions}, as would be expected at two-loop order.

Thus this calculation contains no new information, but rather serves as a consistency check of
our approach.

\subsubsection{The three-point functions}

Using the expressions from \sect{equiv} it is clear that at order $\epsilon^4$, the
three-point functions that appear involve two insertions of $i\bF_{i3}$ and one
operator from $\cS^{(2)}$ in \eqn{exp_ops}.
the three-point function contribution to the operator expansion is
\begin{align}
&\frac{1}{4}\oint ds_1\,ds_2\,ds_3
\bigg(\llangle i\bF_{i3}(s_1) i\bF_{j3}(s_2) iD_{\{k}\bF_{m\}3}\rrangle\epsilon_1^i \epsilon_2^j \epsilon_3^k \epsilon_3^m
+\frac{1}{3}\llangle i\bF_{i3}(s_1) i\bF_{j3}(s_2) iD^m\bF_{m3}\rrangle \epsilon_1^i \epsilon_2^j \epsilon_3^k \epsilon_{3k}
\nonumber\\&\hskip2cm{}
+\llangle i\bF_{i3}(s_1) i\bF_{j3}(s_2) iF_{km}\rrangle \epsilon_1^i \epsilon_2^j \epsilon_3^k \dot\epsilon_3^m
+\llangle i\bF_{i3}(s_1) i\bF_{j3}(s_2) \Phi^1(s_3) \rrangle \epsilon_1^i \epsilon_2^j \dot\epsilon_3^k \dot\epsilon_{3k}
\bigg)\,.
\end{align}
Both $i\bF_{i3}$ and $\Phi^1$ are conformal primaries and applying the constraints of
the residual conformal symmetry, as described in \sect{intro} we expect
\beq
\llangle i\bF_{i3}(s_1) i\bF_{j3}(s_2) \Phi^1(s_3)\rrangle
=\frac{c^0_{\Phi} \, \eta_{ij}}{|s_{12}|^3 |s_{13}| |s_{23}|}
+\cO(\lambda)\,.
\eeq
We emphasize again that the
three-point functions
which
appears at two-loop order in the expectation value of the deformed Wilson loop
are the tree-level coefficients.
We have seen in \sect{equiv} that up to mixing with operators that do not
contribute at this order, $iD^i\bF_{i3}$ is a second level descendant of $\Phi^1$.
Thus we expect
\beq\label{des_con1}
\llangle i\bF_{i3}(s_1) i\bF_{j3}(s_2) iD^k\bF_{k3}(s_3)\rrangle
=-\frac{d^2}{ds_3^2}\llangle i\bF_{i3}(s_1) i\bF_{j3}(s_2) \Phi^1(s_3)\rrangle\,,
\eeq
which serves as another consistency check.

The remaining two operators above, $iF_{ij}$ and $iD_{\{i}\bF_{j\}3}$,
are conformal primaries, one in the $\bf3$ and one in the $\bf5$ or $SO(1,2)$
and the three-point functions involving these primaries have a complicated
tensor structure, but we label the overall prefactors by $c^0_{\bf3}$ and
$c^0_{\bf5}$.

To find these coefficients, we study the triple integrals in the two-loop expression
of the Wilson loop \eqn{gen_2l} and expand the curve to fourth order in $\epsilon$.
In addition to the original triple integral on the first line of \eqn{gen_2l}, there are
boundary terms from integrating the quadruple integrals by parts. As discussed
above at order $\epsilon^2$, the resulting expressions are divergent and careful
regularization is required. After following the procedure outlined there, we find that
that tensor structures of the complicated three-point functions are%
\footnote{We verified the tensor structure of the first one independently from
conformal symmetry, but not of the second one.}
\bal
&\llangle i\bF_{i3}(s_1) i\bF_{j3}(s_2) iF_{km}(s_3)\rrangle
=\frac{c^0_{\bf3} \, \left(\eta_{ik} \eta_{jm} - \eta_{im} \eta_{jk}\right)}{|s_{12}|^2 |s_{13}|^2 |s_{23}|^2}
+\cO(\lambda)\,,
\\
&\llangle i\bF_{i3}(s_1) i\bF_{j3}(s_2) iD_{\{k}\bF_{m\}3}(s_3)\rrangle
=\frac{c^0_{\bf5} \, \left(\eta_{ik} \eta_{jm} + \eta_{im} \eta_{jk} - \frac{2}{3}\eta_{ij} \eta_{km}\right)}
{|s_{12}|^2 |s_{13}|^2 |s_{23}|^2}
+\cO(\lambda)\,,
\eal
The calculation also provides the values of the tree-level structure constants as
\beq
c^0_{\Phi}=-\frac{1}{32\pi^4}\,,
\qquad
c^0_{\bf3}=\frac{1}{16\pi^4}\,,
\qquad
c^0_{\bf5}=\frac{5}{16\pi^4}\,.
\eeq
The calculation also confirms that the descendancy condition \eqn{des_con1} is satisfied.

\subsubsection{Two-point functions}

We now proceed to discuss the two-point functions. Here there are many terms and
we write only those who do not vanish---those between operators in the same multiplets.

The information for the $\Phi^A$ supermultiplet could already be gleaned from the
$O(\epsilon^2)$ analysis in Section~\ref{sec:ep2}. The new information is about the
two-point functions of operators in the $\Phi^1$ supermultiplet.

The terms
in the operator expansion
involving $\Phi^1$ and its conformal descendants
in the operator expansion
are
\bal
\label{Phi1ep4}
\oint ds_1\,ds_2
\bigg(&\frac{1}{72}\llangle iD^k\bF_{k3}(s_1) iD^m\bF_{m3}(s_2)\rrangle
\epsilon_1^i \epsilon_{1i}\epsilon_2^j \epsilon_{2j}
+\frac{1}{8}\llangle \Phi^1(s_1) \Phi^1(s_2) \rrangle
\dot\epsilon_1^i \dot\epsilon_{1i} \dot\epsilon_2^j \dot\epsilon_{2j}
\\&{}
+\frac{1}{12} \llangle \Phi^1(s_1) iD^k\bF_{k3}(s_2) \rrangle
\dot\epsilon_1^i \dot\epsilon_{1i}\epsilon_2^j \epsilon_{2j}
\bigg)\,.
\eal
The information we expect to extract is $a^0_\Phi$ and $\gamma_\Phi$, the tree-level two-point
function coefficient of $\Phi^1$ and its anomalous dimension. Since the latter does
not vanish, the value of $a^1_\Phi$ is scheme dependant, and indeed we could easily
reproduce different values by changing details of our regularization procedures.

The calculation of the three and four-point functions above, which required
integrating \eqn{gen_2l} by parts left many double integrals. In the resulting expressions
we identified the terms proportional to the combinations of $\epsilon_1$ and $\epsilon_2$
in \eqn{Phi1ep4}, leading to the values
\beq
a^0_\Phi=\frac{1}{8\pi^2}\,,
\qquad
\gamma_\Phi=
\frac{1}{4\pi^2}\,.
\eeq
The anomalous dimension matches that found in \cite{alday}, where it was calculated
from Feynman diagrams.

In this supermultiplet we also have the $iF_{ij}$ multiplet, which clearly should have
the same anomalous dimension, but possibly a different two-point function coefficient
\beq
\frac{1}{4}\int_{s_1>s_2}ds_1\,ds_2\,
\llangle iF_{ij}(s_1) iF_{km}(s_2) \rrangle \epsilon_1^i \dot\epsilon_1^j \epsilon_2^k \dot\epsilon_2^m\,.
\eeq
Indeed we find
\beq
a^0_{\bf3}=-\frac{1}{2\pi^2}\,,
\qquad
\gamma_{\bf3}=
\frac{1}{4\pi^2}\,.
\eeq

Similarly, we have
\beq
\int_{s_1>s_2}ds_1\,ds_2\,
\frac{1}{4}\llangle iD_{\{i}\bF_{j\}3}(s_1) iD_{\{k}\bF_{m\}3}(s_2)\rrangle \epsilon_1^i \epsilon_1^j \epsilon_2^k \epsilon_2^m\,,
\eeq
and we find
\beq
a^0_{\bf5}=
\frac{5}{\pi^2}\,,
\qquad
\gamma_{\bf5}=
\frac{1}{4\pi^2}\,.
\eeq

For completeness, we include the terms at this order from the $\Phi^A$ multiplet. Those are
\bal
\label{complete}
&\oint ds_1\,ds_2
\bigg(
\frac{2}{45}\llangle i\bF_{i3}(s_1) iD^n D_{(n} \bF_{m)3}(s_2) \rrangle \eta_{jk}
\epsilon_1^i \epsilon_2^j \epsilon_2^k \epsilon_2^m
\\&{}+
\frac{1}{18} \llangle i\bF_{i3}(s_1) iD_j D^n \bF_{n3}(s_2) \rrangle \eta_{km} \epsilon_1^i \epsilon_2^j \epsilon_2^k \epsilon_2^m
-\frac{1}{6}\llangle i\bF_{i3}(s_1)i\bF_{j3}(s_2)\rrangle
\epsilon_1^i \epsilon_2^j \dot\epsilon_2^k \dot\epsilon_{2k}
\\&{}
+\frac{1}{9}\llangle i\bF_{i3} (s_1) i D^n F_{nm} (s_2) \rrangle
\eta_{jk} \epsilon_1^i \epsilon_2^j \epsilon_2^k \dot\epsilon_2^m
\bigg)\,.
\eal
All of the physical information in the correlation functions above should be
fixed by the descendency relation from the expressions for $\bF_{i3}$ \eqn{aF1}
in Section~\ref{sec:ep2}. When trying to identify these expressions in the Wilson loop 
expansion, we find that the anomalous dimension indeed vanishes as required. 
However, we find extra terms that do not match with the
second order $a^{1}_F$ calculated above. In addition, in cutoff regularization we find 
also the terms
\beq
-\frac{1}{16\pi^4}(\dot\epsilon^1\cdot\dot\epsilon^1)(\dot\epsilon^1\cdot\epsilon^2)
+\frac{1}{16\pi^4}(\epsilon^1\cdot\dot\epsilon^2)(\dot\epsilon^2\cdot\dot\epsilon^2)
+\frac{1}{64\pi^4}(\epsilon^1\cdot\dot\epsilon^1)(\ddot\epsilon^1\cdot\epsilon^2)
-\frac{1}{64\pi^4}(\epsilon^1\cdot\ddot\epsilon^2)(\epsilon^2\cdot\dot\epsilon^2)
\eeq
that do not appear in any of the expression above. 
The coefficients of these unwanted terms (as well as the expression for $a^{1}_F$) 
depend on details of the regularization procedure, the exact choice of cutoffs and
mass terms. This indicates some subtle inconsistencies in our regularization
schemes (or computational errors), as $a^{(1)}_F$ should be well defined. 
Since there were no such
problems in the calculation at order $\epsilon^2$, we expect that the results
in Section~\ref{sec:ep2} including the value of $a^{(1)}_F$ derived there 
are correct (and indeed they agree with \cite{semenoff}).
Likewise all the quantities stated above, the tree-level two-point function
coefficient and anomalous dimensions are robust in our calculation, as are
the structure constants, and we expect them to be correct and scheme 
independent.

\section{Discussion}
\label{sec:discuss}

In this note we studied the properties of operator insertions into the $1/2$ BPS
Wilson loop of $\cN=4$ SYM theory. The expectation value of the Wilson loop
with insertions has the structure of a (defect) CFT and therefore operators can
be assigned a normalization, conformal dimensions and structure constants.

To do this, we first classified operator insertions according to the residual
symmetry group of the space-like Wilson line, $OSp(2,2|4)$. We then
considered the expectation value at one and two-loop order
of smooth curves which are small deformations of the straight line.
Expanding those in the deformation parameter and manipulating the integrals
we found expressions resembling two, three and four-point functions of operators of
different dimensions. We could then match these expressions to particular
operator insertions, via the correspondence between small deformations of
Wilson loops and operator insertions.

Some of the information we found has been known before: The dimension of
$\Phi^A$ and its descendants, which are protected, its one and two-loop
two-point function coefficient, and the one-loop anomalous dimension of $\Phi^1$.
Beyond that we found several more two-point function coefficients and several
structure constants.

It may be possible to use our approach to study operators of higher dimensions
by expanding to higher order in the deformation parameter $\epsilon$. To go to
higher orders in perturbation theory would require general formulae for the
three-loop expectation value of a Wilson loop, which is far beyond current technology.
At that order further issues, like operator mixing which did not effect our calculation,
will arise. An alternative approach to this problem would employ integrability,
which should rather easily provide anomalous dimensions to higher orders
(by analytic perturbative calculation or numerical evaluation at finite coupling).
Further developments would be required to find structure
constants.

In the case of insertion along a Wilson loop with an arbitrary smooth contour, 
the loop breaks conformal invariance completely and one cannot define 
anomalous dimensions anymore. Still, the local analysis is identical, and since 
it is known that loops are renormalized in a multiplicative fashion, with a factor 
for each insertion (and cusp) 
\cite{Brandt:1981kf, Craigie:1980qs, Brandt:1982gz, Dorn:1986dt}, 
those renormalization factors 
can be taken from our analysis of the straight line.

One may also want to study these quantities at strong coupling, via the
$AdS$/CFT correspondence. An efficient algorithm exists to calculate the
expectation value of Wilson loops which are deformations of the $1/2$ BPS
circle (or line) to high orders in $\epsilon$ \cite{dekel}. This is based on
a rather general approach to the problem of minimal surfaces in $AdS_3$
\cite{kruczenski}, so the Wilson loops have to be restricted to
$\bR^2\subset\bR^{1,3}$. This means that the deformations $\epsilon^i$
have to be in a single direction normal to the line, which may lead to some
confusions among different $SO(1,2)$ tensors.

At order $\epsilon^2$ this would reproduce the known large coupling
behavior of the two-point function coefficient of the first insertion $i\bF_{23}$.
At order $\epsilon^4$ there would be again several two, three and four-point
functions which would contribute to the calculation. In contrast to the weak
coupling calculation presented here, the structure of the four-point function
is not restricted to factorize. Rather it can take the form of \eqn{gen-4}
with an arbitrary function $G_{\bF\bF\bF\bF}(u,v)$. While a finite number
of deformations can determine the two and three-point functions, an
infinite number of deformations would be required to fix this. Still, since
$\epsilon$ can be taken to be general, it may be possible to solve for this
function.

At higher order in $\epsilon$ one would encounter five, six and higher
point functions and the problem can get more complicated. Of course, one
may try to ignore these issues, as well as that of the four-point function above,
as knowledge of the two and three-point functions determines a CFT. So one
can disregard the four and higher point function contributions (or match them
as in the last paragraph, or using the OPE.
We leave these issues to the future.

\section*{Acknowledgments}

It is a pleasure to thank Harald Dorn, 
Amit Sever and Kostya Zarembo for illuminating discussions.
We would also like to the organizers and participants of the
``Focus program'' at Humboldt U. Berlin 2016, where parts
of this work were presented.
The work of N.D. is supported by Science \& Technology Facilities Council
via the consolidated grant number ST/J002798/1. The work of A.D.
is supported by the Swedish Research Council (VR) grant 2013-4329.

\bibliography{ref}
\end{document}